\documentclass[aps,prl,twocolumn,showpacs,groupedaddress]{revtex4}
\usepackage{graphics}
\usepackage{graphicx}

\begin{document}


\title{Enhanced Transmission of Light and Particle Waves through Subwavelength Nanoapertures by Far-Field Interference}


\author{S.V. Kukhlevsky}
\affiliation{Department of Physics, University of P\'ecs,
Ifj\'us\'ag u.\ 6, H-7624 P\'ecs, Hungary}


\begin{abstract}

Subwavelength aperture arrays in thin metal films can enable enhanced
transmission of light and matter (atom) waves. The phenomenon relies on resonant excitation and interference of the plasmon or matter waves on the metal surface. We show a new mechanism that could provide a great resonant and nonresonant transmission enhancement of the light or de Broglie particle waves passed through the apertures not by the surface waves, but by the constructive interference of diffracted waves (beams generated by the apertures) at the detector placed in the far-field zone. In contrast to other models, the mechanism depends neither on the nature (light or matter) of the beams (continuous waves or pulses) nor on material and shape of the multiple-beam source (arrays of 1-D and 2-D subwavelength apertures, fibers, dipoles or atoms). The Wood anomalies in transmission spectra of gratings, a long standing problem in optics, follow naturally from the interference properties of our model. The new point is the prediction of the Wood anomaly in a classical Young-type two-source system. The new mechanism could be interpreted as a non-quantum analog of the superradiance emission of a subwavelength ensemble of atoms (the light power and energy scales as the number of light-sources squared, regardless of periodicity) predicted by the well-known Dicke quantum model.
\end{abstract}


\maketitle

%
\maketitle
%
%
The scattering of waves by apertures is one of the basic phenomena
in the wave physics. The most remarkable feature of the light
scattering by subwavelength apertures in a metal screen is
enhancement of the light by excitation of plasmons in the
metal. Since the observation of enhanced transmission of light
through a 2D array of subwavelength metal nanoholes \cite{Ebb},
the phenomenon attracts increasing interest of researchers because
of its potential for applications in nanooptics and nanophotonics
\cite{Barn,Schr,Sobn,Port,Asti,Taka,Lala,Barb,Kukh1,Leze,More,Liu,Garc,Tre,Cao,Sar,Bar,Koer,Miro,Pend,Gome,Mo,Ung,Hua,Ben}. The enhancement of light is a process that can
include resonant excitation and interference of surface plasmons
\cite{Schr,Sobn,Port}, Fabry-Perot-like intraslit modes
\cite{Asti,Taka,Lala,Barb,Kukh1}, and evanescent electromagnetic waves
at the metal surface \cite{Leze}. In the case of a thin screen
whose thickness is too small to support the intraslit resonance,
the extraordinary transmission is caused by the excitation and interference of
plasmons on the metal surface \cite{Schr,Sobn,Port}. Recently, the enhanced transmission through subwavelength apertures by excitation of the 
matter-wave analog of surface plasmons was predicted also for de Broglie matter waves \cite{More}. 
For some experimental conditions, many studies~\cite{Tre,Cao,Sar,Bar,Koer,Miro,Pend,Gome,Mo,Ung,Hua,Ben} indicated an essential role of the surface plasmons in the enhancement of light waves. For an example, the study \cite{Pend} showed that a perfect conductor whose surface is patterned by an array of holes can support surface polaritons, which just mimic a surface plasmon. Nowadays, it is generally accepted~\cite{Liu} that the excitation and interference of surface plasmon-polaritons play a key role in the process of enhancement of light waves in the most of experiments (also, see a recent comprehensive reviews \cite{Garc,Sary}). In the present study~\cite{Ku}, we show a new mechanism that could provide a great resonant and 
nonresonant transmission enhancement of the light or de Broglie particle waves passed through the apertures not by the surface waves, but by the constructive interference of diffracted waves (beams generated by the apertures) at the detector placed in the far-field zone.

The transmission enhancement by the constructive interference of diffracted waves at the detector can be explained in terms of the following theoretical formulation. We first consider the transmission of light through a structure that is similar, but simpler than an array of holes, namely an array of parallel subwavelength-width slits in the metal screen. In some respects, the resonance excitation and interference of surface plasmon-polaritons in these two systems are different from each other~\cite{Po}. The difference, however, is irrelevant from a point of view of our model. Indeed, the excitation of plasmon-polaritons and coupling between the apertures do not affect the principle of the enhancement based on the constructive interference of diffracted waves (beams generated by the independent apertures) at the detector placed in the far-field zone. The resonant excitation of the plasmons or trapped electromagnetic modes, as well as the coupling between apertures could provide just additional, in comparison to our model, enhancement by increasing the power (energy) of each beam. Therefore, our model considers an array of slits, which are completely independent from each other. We also assume, for the sake of simplicity, that the metal is a perfect conductor. Such a metal is described by the classic Drude model for which the plasmon frequency tends towards infinity. The beam produced by each independent slit is found by using the Neerhoff and Mur model, which uses a Green's function formalism for a rigorous numerical solution of Maxwell's equations for a single, isolated slit \cite{Neer,Harr,Betz,Kukh2,Mech,Kukhl}. In the model, the screen placed in vacuum is illuminated by a
normally incident TM-polarized wave with the wavelength
$\lambda=2{\pi}c/\omega=2\pi/k$. The magnetic field of the incident wave
${\vec{H}}(x,y,z,t)=U(x){\exp}(-i(kz+\omega{t})){\vec{e}}_y$ is
supposed to be time harmonic and constant in the $y$ direction. 
The transmission of the slit array is determined by calculating all the light power of the ensemble of beams in the observation plane. To clarify the numerical results, we then present an analytical model, which quantitatively explains the resonant and nonresonant enhancement in the intuitively transparent terms of the constructive interference of diffracted waves (beams generated by the apertures) at the detector placed in the far-field zone. Finally, we show that the mechanism depends neither on the nature (light or matter) of the beams (continuous waves or pulses) nor on material and shape of the multiple-beam source (arrays of 1-D and 2-D subwavelength apertures, fibers, dipoles or atoms).

Let us first investigate the light transmission versus the wavelength by using the
rigorous numerical model. The model considers an ensemble of $M$ waves (beams) 
produced by $M$ independent slits of width $2a$ and period $\Lambda$ in a screen of thickness $b$. 
The transmission of the slit array is determined by calculating all
the light power $P(\lambda)$ radiated by the slits into the far-field
diffraction zone, $x{\in}[-\infty,\infty]$ at the distance
$z\gg{\lambda}$ from the screen. The total per-slit transmission
coefficient, which represents the per-slit enhancement in
transmission achieved by taking a single, isolated slit (beam) and
placing it in an $M$-slit ($M$-beam) array, is then found by using an
equation $T_M(\lambda)=P(\lambda)/MP_1$, where $P_1$ is the power
radiated by a single slit. Figure 1 shows the transmission
coefficient $T_M(\lambda)$, in the spectral region 500-2000 nm,
calculated for the array parameters: $a=100$ nm, $\Lambda=1800$
nm, and $b=5\times 10^{-3}\lambda_{max}$.
\begin{figure}
\begin{center}
\includegraphics[keepaspectratio, width=1\columnwidth]{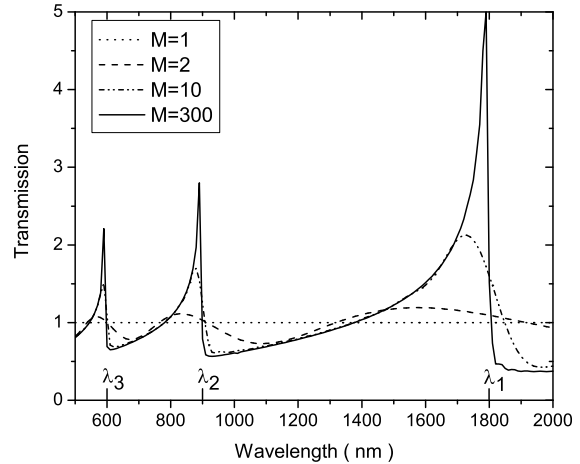}
\end{center}
\caption{The per-slit transmission $T_M(\lambda)$ of an array of
independent slits of period $\Lambda$ in a thin ($b\ll\lambda$) screen versus the wavelength for the different number $M$ of slits. There are three Fabry-Perot like resonances at the wavelengths $\lambda_n{\approx}\Lambda/n$, where $n$=1, 2 and 3.}
\label{fig:Fig1}
\end{figure}
The transmitted power was computed by integrating the total energy
flux at the distance $z$ = 1 mm over the detector region of width
$\Delta{x}$ = 20 mm. The transmission spectra $T_M(\lambda)$ is
shown for different values of $M$. We notice that the spectra
$T_M(\lambda)$ is periodically modulated, as a function of
wavelength, below and above a level defined by the transmission
$T_1(\lambda)=1$ of one isolated slit. As $M$ is increased from 2
to 10, the visibility of the modulation fringes increases
approximately from 0.2 to 0.7. The transmission $T_M$ exhibits the
Fabry-Perot like maxima around wavelengths
$\lambda_{n}=\Lambda/n$. The spectral peaks increase with
increasing the number of slits and reach a saturation
($T_M^{max}\approx5$) in amplitude by $M=300$, at
$\lambda\approx{1800}$ nm. The peak widths and the spectral shifts
of the resonances from the Fabry-Perot wavelengths decrease with
increasing the number $M$ of beams (slits). An analysis of Fig.~1, indicates 
that the power (energy) enhancement and dispersion are the general interference properties of 
the ensemble of beams. Therefore, the enhancement and suppression in the
transmission spectra could be considered as the natural properties also of the periodic array of 
independent subwavelength slits. The spectral peaks are characterized by asymmetric Fano-like
profiles. Such modulations in the transmission spectra are known
as Wood's anomalies. The minima and maxima correspond to Rayleigh
anomalies and Fano resonances, respectively~\cite{Hess}. The Wood anomalies in transmission spectra of gratings, a long standing problem in optics, follow naturally from the interference properties of our model. The new point, in comparison to other models~\cite{Scho,Lalan}), is the prediction of a weak Wood anomaly in a classical Young-type two-source system (see, Fig.~1). 

The above-presented analysis is based on calculation of the energy
flux of a beam array, in which the electromagnetic field of a single beam is evaluated numerically. The
transmission enhancement and dispersion were achieved by taking a single, isolated
slit (beam) and placing it in a slit (beam) array. 
The interference of diffracted waves (beams generated by the slits) at the detector placed in the far-field zone
could be considered as a physical mechanism responsible for the enhancement and dispersion. To clarify the
results of the computer code and gain physical insight into the enhancement
mechanism, we have developed an analytical model, which yields
simple formulas for the electromagnetic field of the beam produced by a single slit. For the field
diffracted by a narrow ($2a\ll\lambda, b\geq0$) slit into the
region $|z|> 2a$, the Neerhoff and Mur model simplifies to an analytical one~\cite{Kukh3}. For the magnetic
$\vec{H}=(0,H_y,0)$ and electric $\vec{E}=(E_x,0,E_z)$ components of the single beam we
found the following analytical expressions:
\begin{eqnarray}
{H_y}(x,z)=i{a}DF_0^{1}(k[x^2+z^2]^{1/2}),
\end{eqnarray}
\begin{eqnarray}
E_{x}(x,z)={{-az}{[x^2+z^2]^{-1/2}}}D
F_1^{1}(k[x^2+z^2]^{1/2}),
\end{eqnarray}
and
\begin{eqnarray}
E_{z}(x,z)={{ax}{[x^2+z^2]^{-1/2}}}D
F_1^{1}(k[x^2+z^2]^{1/2}),
\end{eqnarray}
where
\begin{eqnarray}
\label{sz:D:def} D=4k^{-1}[[\exp(ikb)(aA-k)]^{2}-(aA+k)^2]^{-1}
\end{eqnarray}
and
\begin{eqnarray}
\label{sz:A:def}
A=F_0^{1}(ka)+\frac{\pi}{2}[\bar{F}_{0}(ka)F_1^{1}(ka)
+\bar{F}_{1}(ka)F_0^{1}(ka)].
\end{eqnarray}
Here, $F_1^{1}$, $F_0^{1}$, $\bar{F}_{0}$ and $\bar{F}_{1}$ are
the Hankel and Struve functions, respectively. The beam is
spatially inhomogeneous, in contrast to a common opinion that a
subwavelength aperture diffracts light in all directions uniformly
\cite{Lez}. The electrical and magnetic components of the field produced 
by a periodic array of $M$ independent slits (beams) is given by
$\vec{E}(x,z)=\sum_{m=1}^{M}\vec{E}_{m}(x+m\Lambda,z)$ and
$\vec{H}(x,z)=\sum_{m=1}^{M}\vec{H}_{m}(x+m\Lambda,z)$, where
$\vec{E}_{m}$ and $\vec{H}_{m}$ are the electrical and magnetic components of the $m$-th beam
generated by the respective slit. As an example, Fig.~2(a)
compares the far-field distributions $\vec{E}$ and $\vec{H}$ calculated by the analytical formulae (1-5) to that obtained by the rigorous computer model. We notice that the distributions are undistinguishable.
\begin{figure}
\begin{center}
\includegraphics[keepaspectratio, width=0.9\columnwidth]{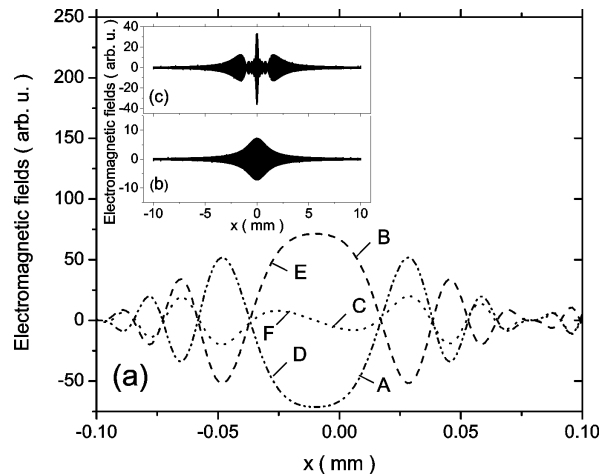}
\end{center}
\caption{The electrical and magnetic components of the field produced 
by an array of $M$ independent slits (beams). (a) The
distributions Re($E_x(x)$) ($A$ and $D$), Re($H_y(x)$) ($B$ and $E$), and
Re($10E_z(x)$) ($C$ and $F$) calculated for $M$ = 10 and $\lambda$
= 1600~nm. The curves $A$, $B$, and $C$: rigorous computer code; curves
$D$, $E$, and $F$: analytical model. (b) Re($E_x(x)$) for $M$=1:
analytical model. (c) Re($E_x(x)$) for $M$=5: analytical model.}
\label{fig:Fig2}
\end{figure}
The field power $P(\vec{E},\vec{H})$ is found by integrating the energy flux
$\vec{S}=\vec{E}\times\vec{H}^*+\vec{E}^*\times\vec{H}$. 
Therefore, the analytical model accurately describes also the coefficient $T_M$ of the system of $M$ independent subwavelength slits (beams). The analytical model not only supports results of our rigorous computer code (Fig. 1), but presents an intuitively transparent explanation (physical mechanism) of the enhancement and suppression in transmission spectra in terms of the constructive or destructive interference of the waves (beams produced by the subwavelength-width sources) at the detector placed in the far-field zone. The array-induced decrease of the central beam divergence by the far-field multiple-beam interference (Figs.~2(b) and 2(c)) is relevant to the beaming light \cite{Mart}, as well as the non-diffractive light and matter beams \cite{Kuk,KukhPRE}. The amplitude of a beam (evanescent spherical-like wave) produced by a single slit rapidly decreases with increasing the distance from the slit (1-3). However, due to the multiple-beam interference mechanism of the enhancement and beaming, the array produces in the far-field zone a propagating wave with low divergence. Such a behavior is in agreement with the Huygens-Fresnel principle, which considers a propagating wave as a superposition of secondary spherical waves. 

We now consider the predictions of our analytical model in light of the key observations published in the literature for the two fundamental systems of
wave optics, the one-slit and two-slit systems. The major
features of the transmission through a single subwavelength slit are the intraslit resonances and the spectral shifts of the resonances from the Fabry-Perot wavelengths \cite{Taka}. In agreement with the predictions
\cite{Taka}, the formula (4) shows that the transmission $T$ =
$P/P_0$ = $(a/k)[$Re$(D)]^{2}+[$Im$(D)]^2$ exhibits Fabry-Perot
like maxima around wavelengths $\lambda_{n}=2b/n$, where $P_0$ is
the power impinging on the slit opening. The enhancement and
spectral shifts are explained by the wavelength dependent terms in
the denominator of Eq. (4). The enhancement
($T(\lambda_1){\approx}b/{\pi}a$ \cite{Kukh3}) is in contrast to the
attenuation predicted by the model \cite{Taka}. 
Although, our model considers a screen of perfect conductivity, polarization charges develop on the metal surface. The surface polaritons do not adhere strictly to traditional surface plasmons. Nevertheless, at the resonant conditions, the system redistributes the electromagnetic energy by the surface polaritons in the intra-slit region and around the screen. Thus, additional energy could be channeled thought the slit in comparison to the energy impinging on the slit opening. The mechanism is somewhat similar to that described in the study \cite{Pend}. This study showed that a perfect conductor whose surface is patterned by an array of holes can support surface polaritons that mimic a surface plasmon in the process of channeling of additional energy into the slit. We considered TM-polarized modes because TE modes are cut off by a thick slit. In the case of a thin screen, TE modes propagate into slit so that magneto-polaritons develop. Because of the symmetry of Maxwell's equations the scattering intensity is formally identical with $\vec{E}$ and $\vec{H}$ swapping roles. Again, the magneto-polaritons could provide channeling of additional energy into the slit. This enhancement mechanism is different from those baced on the constructive interference of the waves (beams produced by the subwavelength-width sources) at the detector placed in the far-field zone. The Young type two-slit (two-beam) configuration is characterized by a sinusoidal modulation of the transmission spectra (for an example, see $T_2(\lambda)$ in Refs. \cite{Scho,Lalan}. The
modulation period is inversely proportional to the slit separation
$\Lambda$. The visibility $V$ of the fringes is of order 0.2,
independently of the slit separation. In our model, the
transmission $T_{2}$ depends on the interference-like cross term ${\int}
[F_1^1(x_1)[iF_0^1(x_2)]^{*}+F_1^1 (x_1)^{*}iF_0^1(x_2)]dx$, where
$x_1=x$ and $x_2=x+\Lambda$. The high-frequency interference-like modulations with
the sideband-frequency $f_s(\Lambda)$
${\approx}f_1(\lambda)+{f_2(\Lambda,\lambda)}{\sim}1/{\Lambda}$
(Figs. 1 and 3) are produced like that in a classic heterodyne system by
mixing two waves having different spatial frequencies, $f_1$ and
$f_2$.
\begin{figure}
\begin{center}
\includegraphics[keepaspectratio, width=0.8\columnwidth]{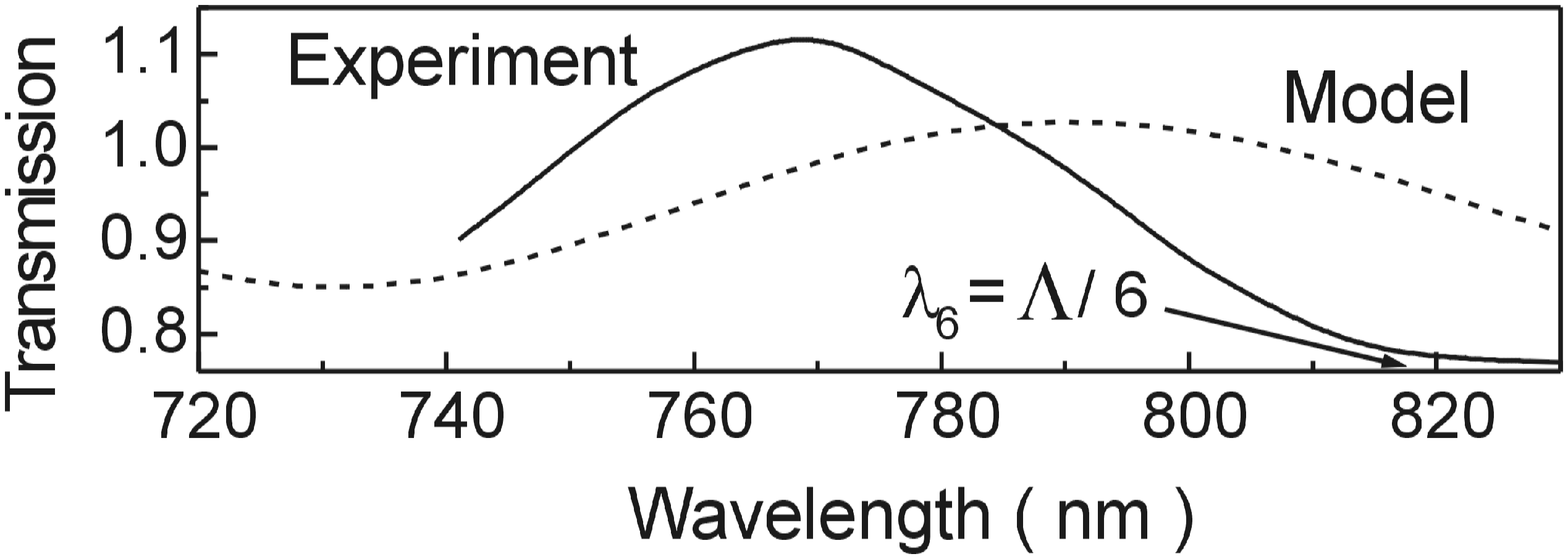}
\end{center}
\caption{The per-slit transmission coefficient $T(\lambda)$ versus
wavelength for the Young type two-slit experiment \cite{Scho}.
Solid curve: experiment; dashed curve: analytical model.
Parameters: $a$ = 100 nm, $\Lambda$ = 4900 nm, and $b$ = 210 nm.}
\label{fig:Fig3}
\end{figure}
Although our model ignores the enhancement by the plasmon-polaritons, its prediction for the transmission
($T_2^{max}{\approx}1.1$),
the visibility ($V{\approx}0.1$) of the fringes and the resonant
wavelengths $\lambda_{n}\approx\Lambda/n$ compare well with the
plasmon-assisted Young's type experiment \cite{Scho} (Fig.~3). 
It should be noted that in the case of $b\geq{\lambda/2}$, the far-field interference resonances at
$\lambda_{n}\approx\Lambda/n$ could be accompanied by the intraslit
polariton resonances at $\lambda_{n}\approx2b/n$. One can easily demonstrate such
behavior by using the analytical formulas (1-5). 
The interference of two beams at the detector is not only the contribution to enhanced transmission. There 
could be enhancement also due to the energy redistribution by the resonant intraslit plasmon-polaritons and/or by the surface waves with resonant coupling through the slits. We stress, however, that the plasmons or trapped electromagnetic modes do not affect the principle of the enhancement based on the constructive interference of diffracted waves (beams generated by the independent subwavelength-width apertures) at the detector placed in the far-field zone. The plasmon-polaritons could provide just additional enhancement by increasing the power (energy) of each beam. This kind of enhancement is of different nature compared to our model, because the model requires nether resonant excitation of the intraslit plasmon-polaritons nor coupling between the slits (see, also Refs.~\cite{Gord,KukhArX,Hua}).

In order to gain physical insight into the mechanism of
plasmonless and polaritonless enhancement in a multiple-slit or multiple-beam ($M{\geq}2$) system, we
now consider the dependence of the transmission $T_M(\lambda)$ on
the slit (beam) separation $\Lambda$. 
\begin{figure}
\begin{center}
\includegraphics[keepaspectratio, width=0.8\columnwidth]{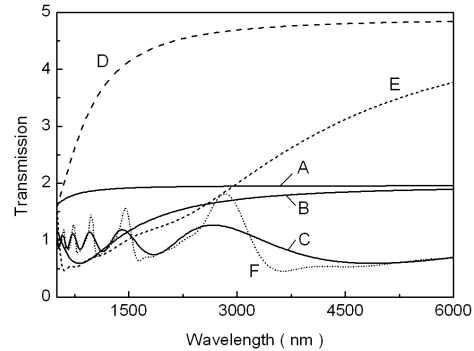}
\end{center}
\caption{The per-slit transmission $T_M(\lambda)$ versus
wavelength for the different values of $\Lambda$ and $M$: (A)
$\Lambda=$ 100 nm, $M=2$; (B) $\Lambda=$ 500 nm, $M=2$; (C)
$\Lambda=$ 3000 nm, $M=2$; (D) $\Lambda=$ 100 nm, $M=5$; (E)
$\Lambda=$ 500 nm, $M=5$; (F) $\Lambda=$ 3000 nm, $M=5$.
Parameters: $a$ = 100 nm and $b$ = 10 nm.  There are two
enhancement regimes at $\Lambda\ll\lambda$ and
$\Lambda{>}\lambda$.} \label{fig:Fig3}
\end{figure}
According to the Van Citter-Zernike coherence theorem, a light
source (even incoherent) of radius $r=M(a+{\Lambda})$ produces a
transversally coherent wave at the distance
$z{\leq}{\pi}Rr/\lambda$ in the region of radius $R$. In the
case of $\Lambda\ll\lambda$, the collective coherent emission of an 
ensemble of slits (beams) generates the coherent electromagnetic
field ($\vec{E}=\sum_{m=1}^{M}\vec{E}_{m}$exp$(i\varphi_m){\approx}M\vec{E}_{1}$exp$(i\varphi)$
and $\vec{H}{\approx}M\vec{H}_{1}$exp$(i\varphi)$) in the far-field zone of the region of radius $R=\infty$. This means that the beams arrive to the detector with the nearly same phases $\varphi_{m}(x)\approx{\varphi{(x)}}$ (see, also Ref.~\cite{Gen}). Consequently, the beams add coherently and the power (energy) of the emitted light scales as the number of beams squared, regardless of periodicity, $P\approx{M^2}P_1$. Thus, the transmission enhancement 
($T_M=P/MP_1$) grows linearly with the number of
slits, $T_M{\sim}M$. For a given value of $M$, in the case of $\Lambda\ll\lambda$, the transmission $T_M(\lambda)$ monotonically (non-resonantly) varies with $\lambda$ (see, Fig. 4). At the appropriate conditions, the transmission can reach the 1000-times nonresonant enhancement
($M={\lambda}z/\pi{R}(a+\Lambda)$). In the case
of $R>{\lambda}z/{\pi}r$ or $\Lambda{>}\lambda$, the beams
arrive to the detector with different phases $\varphi_{m}(x)$. 
Consequently, the power and transmission enhancement grow slowly with the
number of beams (Figs.~1-4). The constructive or destructive
interference of the beams leads respectively to the enhancement or suppression of the transmission amplitudes. Although, the addition of beams is not so efficient, the multiple-beam interference leads to enhancements and resonances (versus wavelength) in the total power transmitted. In such a case, the transmission coefficient $T_M$ exhibits the Fabry-Perot like maxima around the wavelengths $\lambda_{n}=\Lambda/n$. 
We stress again that the constructive or destructive interference of beams at the detector requires nether the resonant excitation of plasmon-polaritons nor the coupling between radiation phases of the slits. The plasmon-polariton effects could provide just additional enhancement by increasing the power (energy) of each beam. Our consideration of the subwavelength gratings is similar in spirit to the dynamical diffraction models \cite{Tre}, Airy-like model \cite{Cao}, and especially to a surface evanescent wave model~\cite{Leze}. In the case of ${\Lambda}>>\lambda $, our model is in agreement with the theories of conventional (non-subwavelength) gratings~\cite{Pet}. 

In the above-presented multiple-beam interference model, we have considered a particular light-source, namely an array of subwavelength metal slits. One can easily demonstrate the interference mediated enhancement and suppression in the transmission and reflection spectra of an arbitrary array of subwavelength-dimension sources of light or de Broglie particle waves by taking into account the interference properties of Young's double-source system. At the risk of belaboring the obvious, we now describe the phenomenon. In the far-field diffraction zone, the radiation from two pinholes
of Young's setup is described by two spherical waves. The light
intensity at the detector is given by $I(\vec
r)=|(E/r_1)exp(ikr_1+\varphi_1)+
(E/r_2)exp(ikr_2+\varphi_2)|^2=I_1+I_2+2(I_1I_2)^{1/2}cos([kr_1+\varphi_1]-[kr_2+\varphi_2])$.
The corresponding energy is
$W=\int{\int}[I_1+I_2+2(I_1I_2)^{1/2}cos([kr_1+\varphi_1]-[kr_2+\varphi_2])]dxdy$.
Here, we use the units ${\Delta}t=1$. In conventional
Young's setup, which contains the pinholes separated by the
distance ${\Lambda}>>\lambda $, the interference cross term
(energy) vanishes. Therefore, the energy is given by $W=\int{\int}(I_1+I_2)dxdy=W_1+W_2=2W_0$,  where
$W_1=W_2=W_0$. In the case of Young's subwavelength system
(${\Lambda}<<\lambda$, correspondingly $r_1=r_2$ for any coordinate $x$ or $y$),
the energy 
$W=W_1+W_2+2\int{\int}(I_1I_2)^{1/2}cos(\varphi_1-\varphi_2)dxdy$.
The first-order correlation term could provide the enhancement or suppression 
of both the intensity and energy of the light field at the detector (see, also Ref.~\cite{Sch,KukhArX}). Indeed, at the phase condition  $\varphi_1-\varphi_2=0$, the energy enhancement is given by $W=4W_0$. In the case of $\varphi_1-\varphi_2=\pi$, the destructive interference of the two waves leads to the zero transmission, $W=0$.
The same phase conditions provide the enhancement or suppression of
transmitted energy by quantum two-source interference (for example, see
formulas 4.A.1-4.A.9 \cite{Scul}). The enhancement or supression by the classic or quantum interferece at the detector depends neither on the nature (light or matter) of the beams (continuous waves or pulses) nor on material and shape of the multiple-beam source (arrays of subwavelength apertures, fibers, dipols or atoms). Due to Babinet's principle, the model predicts the enhancement or supression also in the reflection spectra. 
According to our model, the power (energy) of the light emitted by the subwavelength-dimension ensemble of light sources scales as the number of light-sources squared, regardless of periodicity of the array of sources. Such an effect is not unknown one in the physics. The famous Dicke quantum model of the superradiance emission of a subwavelength ensemble of atoms~\cite{Dick} predicts the same scaling behavior. Therefore, the mechanism described in the present paper could be interpreted as a non-quantum analog of the superradiance emission of a subwavelength ensemble of coherent light-sources. The evident resemblance between our model and the Dicke model also indicates that the interference of waves at the detector could lead to a new effect, namely the enhancements and resonances (versus period of the array) in the total power emitted by the periodic array of quantum oscillators (atoms). A quantum reformulation of our model, which will be presented in the next paper, could also help us to understand better why a quantum entangled state is preserved on passage through a hole array \cite{Alte}.

In conclusion, we have demonstrated a new mechanism that could provide a great resonant and nonresonant transmission enhancement of the light or de Broglie particle waves passed through the subwavelength width slits not by the surface waves, but by the constructive interference of diffracted waves (beams generated by the apertures) at the detector placed in the far-field zone. The model shows that the beams generated by multiple, subwavelength wide slits can have similar phases and can add coherently. If the spacing of the slits smaller than the optical wavelength, then the phases of the multiple beams are nearly the same and beams add coherently (the light power and energy scales as the number of light-sources squared, regardless of periodicity). If the spacing is larger, then the addition is not so  efficient, but still leads to enhancements and resonances (versus wavelength) in the total power transmitted. In contrast to other models, the mechanism depends neither on the nature (light or matter) of the beams (continuous waves or pulses) nor on material and shape of the multiple-beam source (arrays of 1-D and 2-D subwavelength apertures, fibers, dipoles or atoms). The verification of the results by comparison with data published in the literature supports the model predictions. The Wood anomalies in transmission spectra of gratings, a long standing problem in optics, follow naturally from the interference properties of our model. The new point is the prediction of the Wood anomaly in a classical Young-type two-source system. The new mechanism could be interpreted as a non-quantum analog of the superradiance emission of a subwavelength ensemble of atoms (the light power and energy scales as the number of light-sources squared, regardless of periodicity) predicted by the well-known Dicke quantum model. We stress again that the plasmons or trapped electromagnetic modes do not affect the principle of the enhancement based on the classic or quantum interference of diffracted waves (beams generated by the independent subwavelength sources) at the detector placed in the far-field zone. The plasmon-polaritons could provide just additional enhancement by increasing the power (energy) of each beam. The analytical formulas derived in the present study could be useful for experimentalists who develop nanodevices based on transmission and beaming of light or matter waves by subwavelengths apertures.

This study was supported in part by the Framework for European Cooperation in the field of Scientific and Technical Research (COST, Contract No MP0601).

%
\end{document}